\documentclass[12pt,a4paper,final]{iopart}
\usepackage{iopams}
\usepackage{graphicx}
\usepackage[breaklinks=true,colorlinks=true,linkcolor=blue,urlcolor=blue,citecolor=blue]{hyperref}

\usepackage{epsfig}
\usepackage{amssymb}
\usepackage{dcolumn}
\usepackage{bm}

\def\be{\begin{eqnarray}}
\def\ee{\end{eqnarray}}
\def\ba{\begin{array}}
\def\ea{\end{array}}

\begin{document}
\title[Quantum capacitance of Fabry-P\'{e}rot interferometers]{A realistic quantum capacitance model for quantum Hall edge state based Fabry-P\'{e}rot interferometers}
\author{O. Kilicoglu}
\address{Isik University, Department of Physics, 34980 Istanbul, Turkey}
\author{D. Eksi}
\address{Istanbul Yeni Yuzyil University, Vocational School of Health Services, 34010 Istanbul, Turkey}
\author{A. Siddiki}
\address{Physics Department, Faculty of Letters and Sciences, Mimar Sinan Fine Arts University, 34380 Sisli, Istanbul, Turkey}

\ead{afifsiddiki@gmail.com}
\vspace{10pt}
\begin{indented}
\item[]July 2016
\end{indented}
\begin{abstract}

In this work, the classical and the quantum capacitances are calculated for a Fabry-P\'{e}rot interferometer operating in the integer quantized Hall regime. We first consider a rotationally symmetric electrostatic confinement potential and obtain the widths and the spatial distribution of the insulating (incompressible) circular strips using a charge density profile stemming from self-consistent calculations. Modelling the electrical circuit of capacitors composed of metallic gates and incompressible/compressible strips, we investigate the conditions to observe Aharonov-Bohm (quantum mechanical phase dependent) and Coulomb Blockade (capacitive coupling dependent) effects reflected in conductance oscillations. In a last step, we solve the Schr\"odinger and the Poisson equations self-consistently in a numerical manner taking into account realistic experimental geometries. We find that, describing the conductance oscillations either by Aharanov-Bohm or Coulomb Blockade strongly depends on sample properties also other than size, therefore, determining the origin of these oscillations requires further experimental and theoretical investigation.
\end{abstract}

\section{Introduction}
The charge transport measurements performed at quantized Hall effect (QHE) based particle and quasi-particle interferometers provide information on electronic and statistical
properties of the particles \cite{Halperin:2011,marcus:01,Goldman:2005,Heiblum:2006,Rosenow:2009}. In these experiments conductance through a quantum dot (QD) is measured as a function of external magnetic field $B$ or the gate potential(s) $V_g$ defining the electrostatic confinement. Interestingly, conductance presents oscillations as a function of both $B$ and $V_g$. The peak-to-peak periodicity of the oscillations, $\Delta B$ and $\Delta V_g$ respectively, strongly depend on various properties of the devices. On one hand,
the side gate (SG) defined Fabry-P\'{e}rot type interferometers with large interference
areas $(A > 5  ~\mu$m$^2)$ present usual Aharonov-Bohm (AB) periodicity as a function
of the external magnetic field, i.e. the number of enclosed magnetic flux increase
linearly with $B$. Hence the $\Delta B$ periodicity is constant in $B$. Meanwhile, if a voltage is applied to the top-gate (TG) or the voltage on SG is changed, $\Delta V_g$ varies inverse linearly with $B$. On the other hand, small samples
$(A < 3  ~\mu$m$^2)$ show an opposite behavior. In this situation, $\Delta B$ varies linearly with $B$, i.e. inversely proportional to filling factor $\nu$, and $\Delta V_g$ is independent of the $B$ field. In mainstream theoretical works, this behavior is interpreted as a result of charging effects and therefore is called the Coulomb blockade or Coulomb dominated (CD) regime \cite{Halperin:2011,marcus:01,Rosenow:2009}. However, other interpretations are also available in the literature, based on interactions and time-dependent calculations \cite{Goldman:2005,Salman:2013,Cicek:2012}.

The importance of charging effects on the AB interference has already been discussed some two decades ago by Beenakker within the single particle approximation \cite{Beenakker:1991}. They found that, the conductance oscillation period as a function of magnetic field $\Delta B$ is modified by the charging effects, namely by the capacitance $C$ of the QD. It is shown that, the Coulomb blockade of the Aharonov-Bohm effect occurs, whenever charging energy $e^2/C$ becomes comparable or larger than the energy separation of the magnetic field quantized levels. Later, Evans and co-workers calculated the geometrical capacitance of a similar system also taking into account the direct electron-electron interaction, namely they considered the formation of metal-like compressible and insulator-like incompressible strips \cite{Evans:1993}. They elucidated the experimental findings of McEuen et al, where magneto-transport through a relatively small QD (lithographic size of 500 nm x 750 nm, which confines less than 100 electrons) is investigated \cite{McEuen:1992}. In the experimental work, peaks in the conductance were reported and are attributed to charging of the metallic-like compressible island.

Recently, there has been theoretical and experimental reports which essentially emphasises the importance of so-called ``Quantum capacitance". The classical (or equivalently called geometrical) capacitance takes into account only the direct Coulomb interaction and a homogeneous dielectric constant $\varepsilon$. Whereas, in calculating the quantum capacitance $C_{\rm q}$ one should also take into account density of states (DOS), Pauli exclusion principle and correlation effects. As we will summarise below, its essence relies on the fact that the quantum capacitance is just proportional to thermodynamical density of states $D_{\rm T}(\mu)$ (TDOS) at the chemical potential $\mu$ \cite{Gerhardts:2005}. Therefore, systems having a gap at Fermi energy $E_{\rm F}$ in the limit of zero temperature, i.e. $D_{\rm T\rightarrow 0}(E_{\rm F})=0$, the quantum capacitance also vanishes. Since the geometrical $C_{\rm geo}$ and the quantum capacitances are added in series the total capacitance also reads to zero. Hence, to understand capacitive effects (like charging) at gapped systems it is important to investigate the contribution of quantum capacitance, which becomes the dominant counter-part at low temperatures and high $B$ fields.

We organise our paper as follows, first we briefly discuss how the dielectric function is modified due to dimensionality affected DOS considering a two dimensional electron system (2DES). There we also touch the situation where the 2DES is subject to high magnetic fields. Next, we consider a toy model to calculate the capacitance of a rotationally symmetric QD to find the $B$ intervals where the total capacitance is dominated either by the geometric or quantum capacitances, following Beenaker. Afterwards, we show that the oscillation period $\Delta B$ observed at the conduction decreases with increasing field strength only when the geometrical capacitance is taken into account. In contrast, we observe that the period strongly depends on the field strength if one takes into account the quantum capacitance. Our discussion is followed by a Section, where we compare our results which now include quantum capacitances with a previous similar work by Evans et al. There we show that inclusion of quantum effects strongly alter the oscillations at the conductance also in $\Delta V_g$. In Section \ref{sec:cap_halperin}, we calculate the total capacitance of a realistic device and seek for the parameter regimes where the conductance oscillations can be determined either by charging effects (CD) or by interference (AB) effects. In this Section, we first consider only the electrostatic effects and then include self-consistent calculations. We find that although it is possible to observe CD oscillations in real experiments, it is strongly constrained by the sample parameters.

\section{Geometrical and Quantum Capacitances of a homogeneous 2DEG}
In general, capacitance is a measure of energy to be paid to charge a device which is composed of two {\it metals} separated by an {\it insultor} \cite{Serway}. Here, of course we used the words metal and insulator in a hand-waving way. The energy necessary to charge such a device is given by $E_C=Q.V=Q^2.C$, where $V$ is the potential difference between the metals, $Q$ the charge and $C$ the capacitance. Considering the simplest classical device in three dimensions, two metal plates perpendicular to each other and separated by an insulator, the capacitance is given by the area of the metal plate $A$, the distance between metal plates and the dielectric {\it constant} $\kappa$ of the insulator,
\be C=\kappa \frac{A}{d}. \ee
However, as we will briefly discuss below, all our arguments above impose couple of assumptions which are not valid in low-dimensions also including quantum mechanical interactions beyond the direct Coulomb interaction.

First, it is useful to check the common assumptions imposed on the dielectric {\it function} $\varepsilon(\vec r,t)$ \cite{Kittel}. Here, $\vec r$ determines the spatial coordinate and $t$ is the time. Equivalently the dielectric function can be defined by its Fourier adjuncts, $\vec k$ and $\omega$, which are momentum and frequency respectively. The dielectric function can be regarded as the response of a material subject to an external electric field $\vec{E}(\vec r,t)$. Now lets first assume that our material is simultaneously responding to the external field, hence assume that the dielectric function is time (or equally frequency) independent, i.e $\varepsilon(\vec k,\omega)=\varepsilon(\vec k,0)$. Assuming a spatial homogeneity results in a constant dielectric function $\kappa$. To make the connection between the dielectric function and energy dispersion of the system we now lift the assumption on spatial homogeneity \cite{Ashcroft}. Then, it is easy to show that the response of the material to the external field is described by
\be \phi(\vec k)=\frac{1}{\varepsilon(\vec k)}\phi^{\rm ext}(\vec k). \ee
Here we implicitly made the assumption that, the screened potential $\phi(\vec k)$ and the external potential $\phi^{\rm ext}(\vec k)$ are linear. Then, a material can be called metal in the limit $\varepsilon(\vec k)\rightarrow \infty$ ($k\rightarrow 0$). We also assumed that the field is uniform and the potential varies slowly on the scale of the particle separation. Under the assumption of uniformity an insulator can be defined as a material where its dielectric function is finite and small compared to a metal. Now the problem is reduced to find the exact form of $\varepsilon(\vec k)$ including dimensional and quantum effects. In the following we will confine ourself to the lowest order mean field approximation to describe quantum mechanical corrections to $\varepsilon(\vec k)$, namely the Thomas-Fermi approximation.

If one solves the related time-independent Schr\"odinger equation and keep the assumption of slowly varying potential the momentum dependent eigenenergies are described as,
\be \epsilon(\vec k)=\frac{\hbar^2 \vec k^2}{2m}-e\phi(r),\ee
where $m$ is the bare electron mass and $e$ is the charge of an electron. One step further in including quantum mechanical effects is to calculate the position dependent electron number density utilising the Fermi-Dirac distribution so that,
\be n_{\rm el}(\vec r)=\int \frac{d \vec k}{4\pi^3} \frac{1}{\exp[\xi] +1}, {\label{TFA_density}}\ee
where $\xi=\beta[\epsilon(\vec k)-\mu]$ comprises the thermal energy $1/\beta=k_{\rm B}T$, where $T$ is temperature and $k_{\rm B}$ is Boltzmann constant. Then the dielectric function that includes energy dispersion and also the Fermi-Dirac statistics can be written as,
\be \varepsilon(\vec k)=1+\frac{4\pi e^2}{k^2}\frac{\partial n_0}{\partial \mu}, \ee
where $n_0$ is the equilibrium density without the external field. Namely, the Thomas-Fermi dielectric function.

The above equation includes two important information about the system at hand, first in the $k\rightarrow 0$ limit it describes a metal properly and, second via $\frac{\partial n_0}{\partial \mu}$ the effects arise from dimensionality and spin degree of freedom are explicitly included. Note that, the thermodynamic of states TDOS $D_{T}(\mu,T)$ equals to $\frac{\partial n_0}{\partial \mu}$ and describes how the levels are occupied at a given temperature and number of particles. It is also useful to define a wave vector $k_0=\sqrt{4\pi e^2\frac{\partial n_0}{\partial \mu}}$, which will be helpful in defining a length scale to check the validity of our assumptions, below.
Then, the dielectric function reads \cite{Gerhardts:2003},
\be \varepsilon(\vec k)=1+\frac{k_0^2}{k^2}.\ee
Given the definition of capacitance in general, it is now possible to define a quantum mechanically corrected counter part, which now also includes the statistical properties of the system,
\be C_q=e^2D_T(\mu,T),\ee
and is defined per area. Then the total capacitance reads,
\be \label{eq:totalC}1/C_T=1/C_{geo}+1/C_q \ee

It is important to note that, the dielectric function is now a thermodynamic quantity. Namely, it is not only a function of material properties but also temperature, statistics of the particles and how the quantum mechanical states are occupied. Thus, it is indispensable to take into account the limitations of the above expression. One can assign a special role to the wave vector $k_0=2\pi/\lambda_T$, where $\lambda_T$ is a thermal quantity, thermal wavelength. In the limit of zero temperature, $1/k_0$ is strongly related with the Fermi wavelength, which essentially limits validity of above screening argument by the number of particles involved, when utilised. Therefore, if one uses the dielectric function, as a mean field quantity one should also take into account whether if it is valid when the number of particles is sufficiently low. Hence, we will keep our eye on the validity regimes of our assumptions, when considering ``small" ensembles  such as quantum dots and narrow incompressible strips.

\subsection{Homogeneous 2DEG in the absence of an external $B$ field}
Now we are equipped with the minimally corrected dielectric function which comprises also the necessary quantum mechanical and statistical information. In the beginning of this subsection, we will just refer to the well known text book results that describe the TDOS of a homogeneous 2DEG and connect quantum capacitance with the dielectric function.

It is well known that, if one can neglect the boundary effects and also the finiteness of the particle number density (i.e. in the thermodynamical limit) one can write the DOS of a 2DES as,
\be D(E)=D_0= \frac{m}{(\pi\hbar^2)}, \ee
here $m$ is the bare mass of an electron in vacuum, and will be replaced by an effective mass $m^*$ when considering a semi-conductor heterostructure (homogeneous insulator) later. In the limit of $T\rightarrow0$ and utilising the fact that $E_F=\mu$ in 2D Eq.~(\ref{TFA_density}), reduces to
\be n_{\rm el}(x,y)=D_0[E_F-e\phi(x,y)]\Theta[E_F-e\phi(x,y)], \ee
which is a linear relation if, $e\phi(x,y)<E_F$, satisfying our previous assumption.

Recall that, the dielectric function can be expressed in terms of TDOS and dielectric constant of the material $\kappa$ in the limit of zero temperature, namely,
\be \varepsilon(\vec k)=1+ \frac{2\pi e^2D_0}{\kappa|\vec k |}=1+\frac{2}{a^*_B|\vec{k}|},\ee
where $a^*_B=\kappa\hbar^2/(me^2)$ is the effective Bohr radius (For GaAs $a^*_b=9.81$ nm).

Hence, we can write a relation between capacitance per area and the TDOS of a homogeneous unbounded 2DES via dielectric function as,
\be C_q=e^2D_0. \ee
\begin{figure}
\hspace{2.15cm}
\includegraphics[width=0.9\columnwidth]{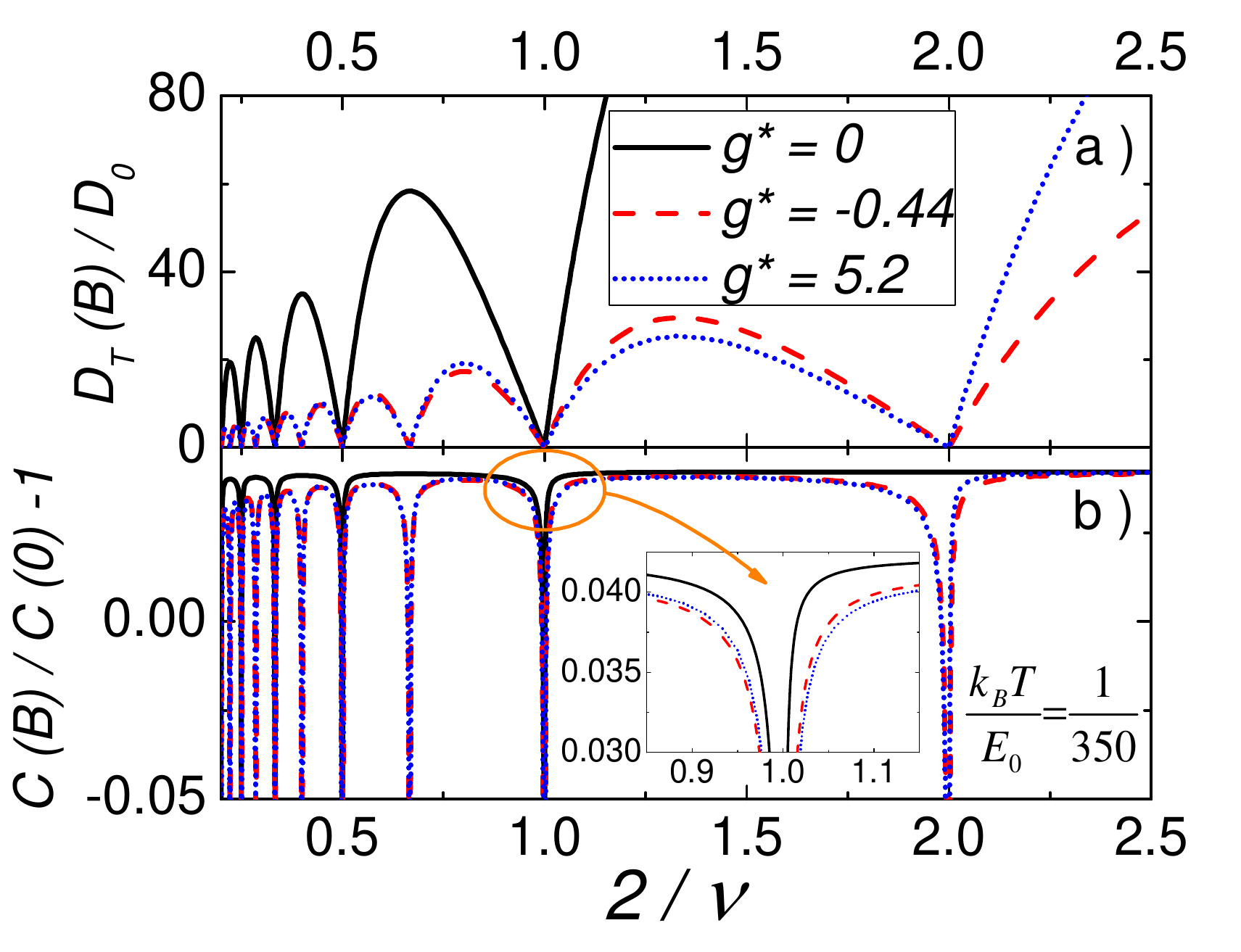}
\caption{(color online)  \label{fig1} (a) The Thermodynamic density of states (TDOSs) calculated according to Eq.(\ref{TermoDynamic_DensityOfState}) considering different $g^*$ values and assuming a homogeneous 2DEG, for increasing $B$ ($=2/\nu$) (b) The total capacitances which are obtained from Eq.(\ref{capacitance_geo_qua}), as a function of the filling factor. Strong dips occur at integer filling factors, since TDOS approaches zero. Note that for $g^*=0$ such behaviour occurs only at even integers. The inset emphasises the considerable difference between Zeeman split and non-split cases.
}
\end{figure}
At this point, we would like to remind the reader the close relation between \emph{compressibility} and \emph{capacitance}, both being thermodynamical quantities. Recall that, capacitance is nothing but the energy required to add an additional charge (per area), which is given by the TDOS. As a well known text book result, at finite temperature, the compressibility is given by:
\be \Gamma=n_{\rm el}^{-2}D_T(\mu). \ee
One sees that if the TDOS vanishes both the capacitance and compressibility vanishes. In other words, once TDOS becomes zero, the system under investigation becomes incompressible and capacitance reads zero. Such a situation can be obtained, if the system at hand has gap at the Fermi energy. An external $B$ field applied perpendicularly to a 2DES provides this opportunity, which we are going to discuss next.

\subsection{Homogeneous 2DEG in the presence of a finite external $B$ field}
Once the relation between the TDOS and capacitance is obtained it is almost straight forward to obtain the magnetic field dependent quantum capacitance per area for a homogeneous 2DES subject to homogeneous perpendicular $B$ field as,
\begin{equation}\label{capacitance_geo_qua}
C_q(B)=e^2D_{\rm T}(B),
\end{equation}
where $D_{\rm T}(B)$ is the TDOS of a homogeneous electron system. Since, the experiments are performed at high mobility samples it is reasonable to assume that the collision broadening of the Landau levels is negligible, hence, the TDOS is given by,
\begin{equation}\label{TermoDynamic_DensityOfState}
D_T(B)={\frac{g_s}{2\pi\ell^2_B}}\sum_{n=0} ^{\infty} \frac{\beta}{4\cosh^2(\beta [E_n - \mu]/2)}.
\end{equation}
Here, $g_s$ is a pre-factor determining the spin degeneracy, $E_n=\hbar\omega_c(n+1/2)$ is the cyclotron energy ($\omega_c=\frac{eB}{m^*}$), and $n$ indexes the spin degenerate Landau levels.  The area of the available states is determined by the magnetic length $\ell_B(=\sqrt{\hbar/eB})$. In a later step we will also take into account Zeeman splitting in an effective field approach. However, we will not consider correlation effects  while we are mainly interested in the integer quantised Hall effect, which is believed to be a single particle effect.

One should make it clear that, disorder is an important parameter from the experimental point of view. As mentioned above, the experiments presenting interference effects are conducted on very high quality samples, which has mean free path much much longer than the size of the quantum dot. Since we only consider the oscillations emanating from the dot, it is we think that the assumption of Dirac like DOS is acceptable, also given the fact that TDOS is also broadened due to temperature. On the other hand, long-range potential fluctuations arising from disorder are at the order of microns. Therefore, it is realistic to ignore the contribution from the disorder both for the DOS broadening and potential fluctuation point of view.

It is useful to introduce a dimensionless parameter, which essentially counts the number of Landau levels below $E_F$, the filling factor
\be \nu=g_s\pi \ell_B^2\bar{n}_{\rm el},\ee where $\bar{n}_{\rm el}$ is the average electron number density. The filling factor is an integer when $E_F$ falls in between two following energy levels. The corresponding thermodynamic quantity is then defined as:
\be \nu(\mu,T)=g_s\sum_{n=0}^{\infty}f(E_n;\mu,T) .\ee

Fig.~\ref{fig1}a, plots the magnetic field dependent TDOS as a function of inverse filling factor (i.e. $\propto B$) at low temperatures. If Zeeman splitting is neglected ($g^*=0$, solid line) one observes zeros only at even integer filling factors, whereas if one also considers Zeeman splitting with the bulk Land\'e $g^*(=-0.44$, broken lines)  or the exchange enhanced $g^*(=-5.2$, dotted lines), one observes additional zeros also at odd filling factors. However, the difference in TDOS between the bulk and exchange enhanced is rather small and is observed only at non-integer filling factors. In Fig.~\ref{fig1}b, we show the corresponding total capacitance calculated from Eq.~(\ref{eq:totalC}), where the geometric capacitance is assumed to be $e^2D_0/23.6$ and at the default temperature $kT/E_0=1/350$ ($E_0=n_0/D_0$ and $T\sim 200-500$ mK). Following the variations at the magnetic field dependency of the TDOS, one observes strong oscillations at the total capacitance. Moreover, it is clearly seen that the quantum capacitance is the dominating term and total capacitance diverges to infinity at the integer filling factors in the $T\rightarrow 0$ limit. In this situation the compressibility becomes zero, hence 2DES becomes incompressible. Of course, at finite temperature the compressibility of the system increases exponentially. However, in the presence of disorder and with localised states at the tails of Landau levels this exponential increase is limited by the number of available localised states.

In the above discussion, we considered an unconfined (homogeneous) 2DES and calculated the total capacitance. We observed that, the quantum counter-part is the dominating term, when the filling factor is an integer. This result, in fact is expected since once all the levels below the Fermi energy are occupied the energy required to add a particle to the system becomes relatively large.

Next, we will utilise the local version of the above formalism considering a QD which is defined by electrostatics and is capacitively coupled to metallic gates. There we will also assume that the confinement potential is slowly varying at the length scale of wave width ($\propto \ell_B$), namely where TFA is still valid.
\begin{figure}
\hspace{2.15cm}
\includegraphics[width=0.9\columnwidth]{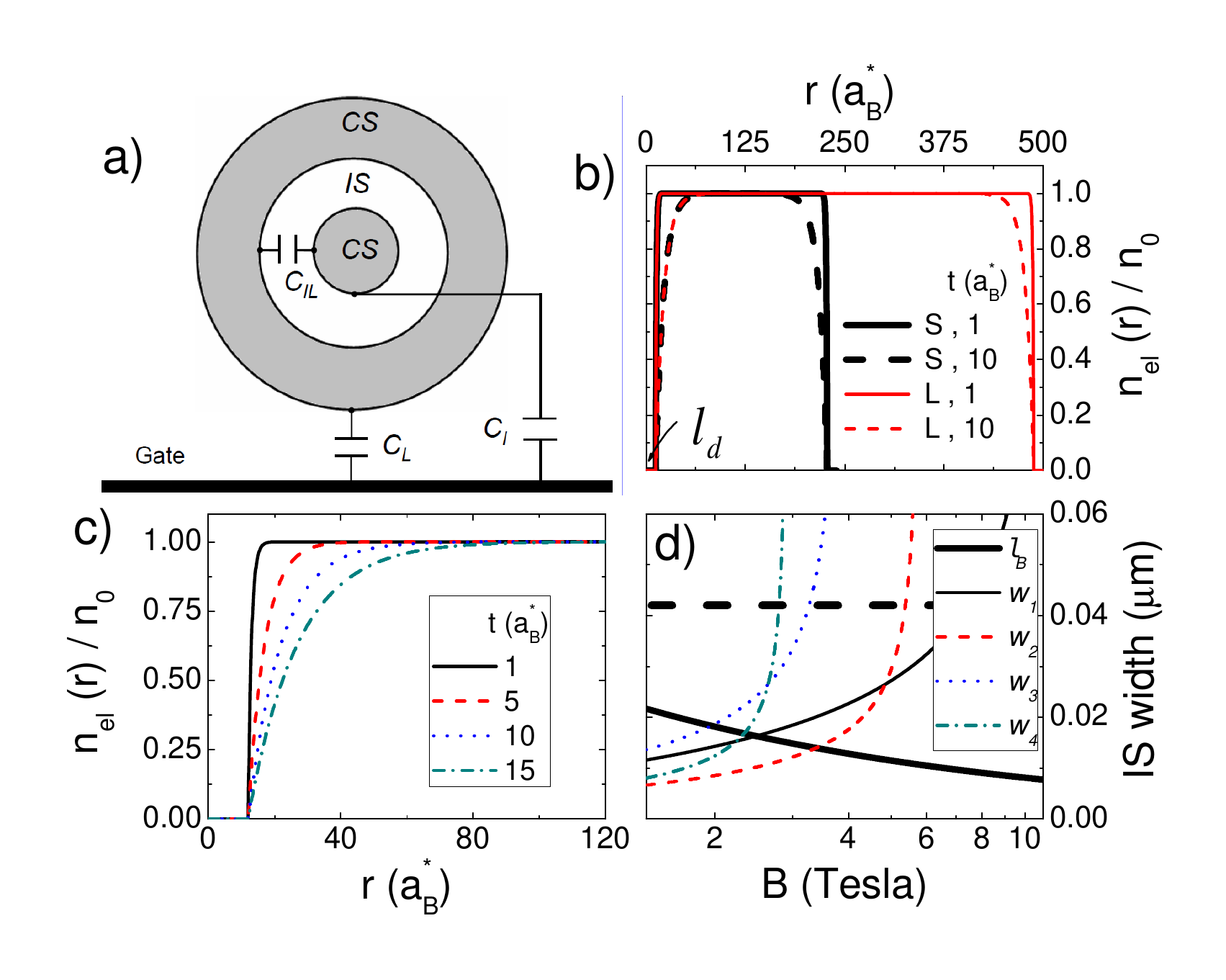}
\caption{\label{fig2} \label{afifhoca_density_ISwidth_capacitances} (a) The schematic presentation of incompressible and compressible strips in the presence of a perpendicular magnetic field, considering a rotationally symmetric QD. (b) Variation of the electron distribution calculated from eq. (\ref{afifhoca_density}), regarding two different sample widths and steepness values. (c) Density distribution as a function of position considering various steepness. (d) The widths of the incompressible strips taking into account the Zeeman splitting (with the parameters extracted from experiments, $g^*=-0.44$, $t=10 a^*_B$, $\ell_d=10a^*_B$ and $a^*_B\approx10$ nm, for GaAs). The bulk electron density is set to be $n_0=2.8\times10^{11}$ cm$^{-2}$, similar to experimental values.
}
\end{figure}
\section{Toy model\label{sec:toy_model}}
In this Section we will calculate both the geometrical and quantum capacitances of a toy FPI operating under integer quantised Hall conditions, within a semi-classical Hartree type approximation. In principle, both the classical and quantum capacitances can be calculated numerically, either by direct diagonalisation or by density functional methods. However, the real QD structures utilised to measure conductance oscillations are relatively large, hence, the number of particles confined to these systems usually exceed thousands. Therefore, both of the techniques are unable to span the related Hilbert space of electrons or quasi-particles, i.e. Khom-Sham particles. Here, we employ a rather simpler method based on a Hartree-type mean field approximation, namely the Thomas-Fermi-Dirac approximation (TFDA) \cite{Bilgec:2010}. The assumptions of TFDA can be summarised as follows: First, the total potential $\phi(r)$, which is composed of confinement and interaction potentials, varies slowly on the quantum mechanical length scales, such as the magnetic length. Hence, the electron wave functions can be replaced by Dirac-delta functions and the corresponding energy eigenvalues are given by \cite{Goldman-spin:2007}
\begin{equation}
E_{n,j}=E_n+g^*\mu_B B S_z+\phi(r_j),
\end{equation}
where $S_z$ is the spin index with $\pm1/2$ and $r_j=(2m^*\hbar/eB)^{1/2}$ is the radius of the drifting cyclotron radius encircling $j$ flux quanta and $m^*=0.067~m_e$, for GaAs, where $m_e$ is the bare electron mass at rest. Second, the spin effects such as Zeeman splitting and exchange potential are taken into account within the Dirac approximation, namely the exchange potential is obtained from density functional approximation where the density is calculated within the Thomas-Fermi approximation. By this approach we can simply replace the bulk Land\'e $g$ factor of the material, by an exchange enhanced effective factor $g^*$ and determine the effective Zeeman gap. It is worthy to note that, our approximation does not include correlations effects. However, since we are only dealing with the integer quantised Hall effect and the correlation effects are thought to be suppressed, higher order many-body effects will be neglected.

To calculate both the geometric and quantum capacitances, one essentially needs to know the spatial and magnetic field dependency of the compressible/incompressible regions \cite{Chklovskii:1992}. Here, we will follow the work by Chklovskii et al, where direct the Coulomb interaction between electrons and quantising perpendicular magnetic field $B$ yield formation of compressible and incompressible strips. In their work, it is analytically shown that once the electro-statically confined 2DES is subject to strong $B$ fields, the electrostatic stability condition results in formation of co-existing electronic regions with different and peculiar screening properties. It is proposed that, starting from the edges of the system, there exists an electron depleted region due to the repulsive force generated by the side gates that define the confinement. The length of the depleted region $\ell_d$ is determined by the potential applied to the metallic gates. This depleted region is followed by a metal-like region, where the Fermi energy is pinned to the lowest Landau level with high DOS. In this region, electron density varies spatially, however, the total potential is (almost) constant due to good screening, similar to a metal, and is called the compressible region. Proceeding to inner parts of the sample, one observes a region where the Fermi energy falls in between two Landau levels, hence, due to vanishing DOS at the Fermi energy, the 2DES becomes insulator-like (locally) and this region is called incompressible. Fig.~\ref{fig2}a, depicts a schematic presentation of such a system together with the relevant capacitances which are denoted by $C_I$, $C_L$ and $C_{IL}$. These capacitances will be calculated considering both geometrical and quantum mechanical effects, in the following subsections.
\subsection{Electron density profile, incompressible strips and capacitances}
To obtain the magnetic field and confinement potential dependent spatial distribution of the compressible (CS) and incompressible strips (IS) one has to first obtain the charge density distribution. In the original work of Chklovskii et al\cite{Chklovskii:1992}, the confinement potential is generated by in-plane metallic gates and donors, which essentially yields a smooth electron density at the edges of the sample. However, self-consistent calculations both in 2D and 3D show that the electron density is rather steep than that of the analytical ones. The deviation is a result of the assumed unrealistic geometry (i.e. in-plane gates and charges) together with the assumption of infinite DOS of the electronic system as if it was a perfect metal ($\varepsilon(\vec k)\rightarrow \infty$). The self-consistent calculations suggest that the electron density distribution can be described by \cite{Ali:2013},
\begin{equation}\label{afifhoca_density}
n(\vec r)=n_0(1-e^{[-(\vec r-\ell_d)/t]}),
\end{equation}
where $\vec r$ is the radial coordinate and $t$ determines the electron poor region in front of the gates fixing the electron density gradient which vanishes for $|r|>R$ and $n_0$ is the bulk electron density. Fig.~\ref{fig2}b depicts the calculated density distribution as a function of radial coordinate considering two sample sizes and two edge potential steepness. Our choice of these parameters are based on the experimental realisation of the samples: In typical experiments two different sizes are considered, meanwhile edges are defined ether by gates (smooth confinement) or by chemical etching (sharp confinement).

We observe that, for the small sample with radius $R\sim 1.2~\mu$m and steep edge $t=1~a_B^*$, the electron density reaches its bulk value rather close to the boundary. Meanwhile, for a smoother edge profile ($t=10~a_B^*$) the bulk value is reached only for a narrow radial interval. For the large sample with radius $R\sim 2.5~\mu$m, the difference between steep and smooth edge profiles is rather small, in the sense that the bulk electron density covers most of the sample and edge effects can be neglected in a first order approximation. Fig.~\ref{fig2}c, provides additional density profiles where the edge potential varies from relatively steep to relatively smooth profile. In a previous work, Salman et. al, reported that the samples defined by chemical etching present steep profiles at the edges, whereas, in-plane gate defined samples show a smoother profile, by 3D self-consistent calculations. However, the exact shape of the edge density profile strongly depends on the couple of parameters, such as the distance of the 2DES and the dopants from the surface, the etching depth, amount of surface charges, area of the top and length of the side gates etc. Therefore, each sample may have a different edge profile, hence, one should perform numerical simulations to obtain a realistic density distribution. In any case, as a rule of thumb we will consider small $t<3$ values to mimic etched and larger $t$ values for gate defined samples. A detailed analysis of various edge profiles can be found in Salman paper.

In the next step one has to obtain the distribution of insulator-like (incompressible) strips to calculate the geometrical capacitances as a function of both $B$ and steepness $t$.
\begin{figure}
\hspace{2.2cm}
\includegraphics[width=.84\columnwidth]{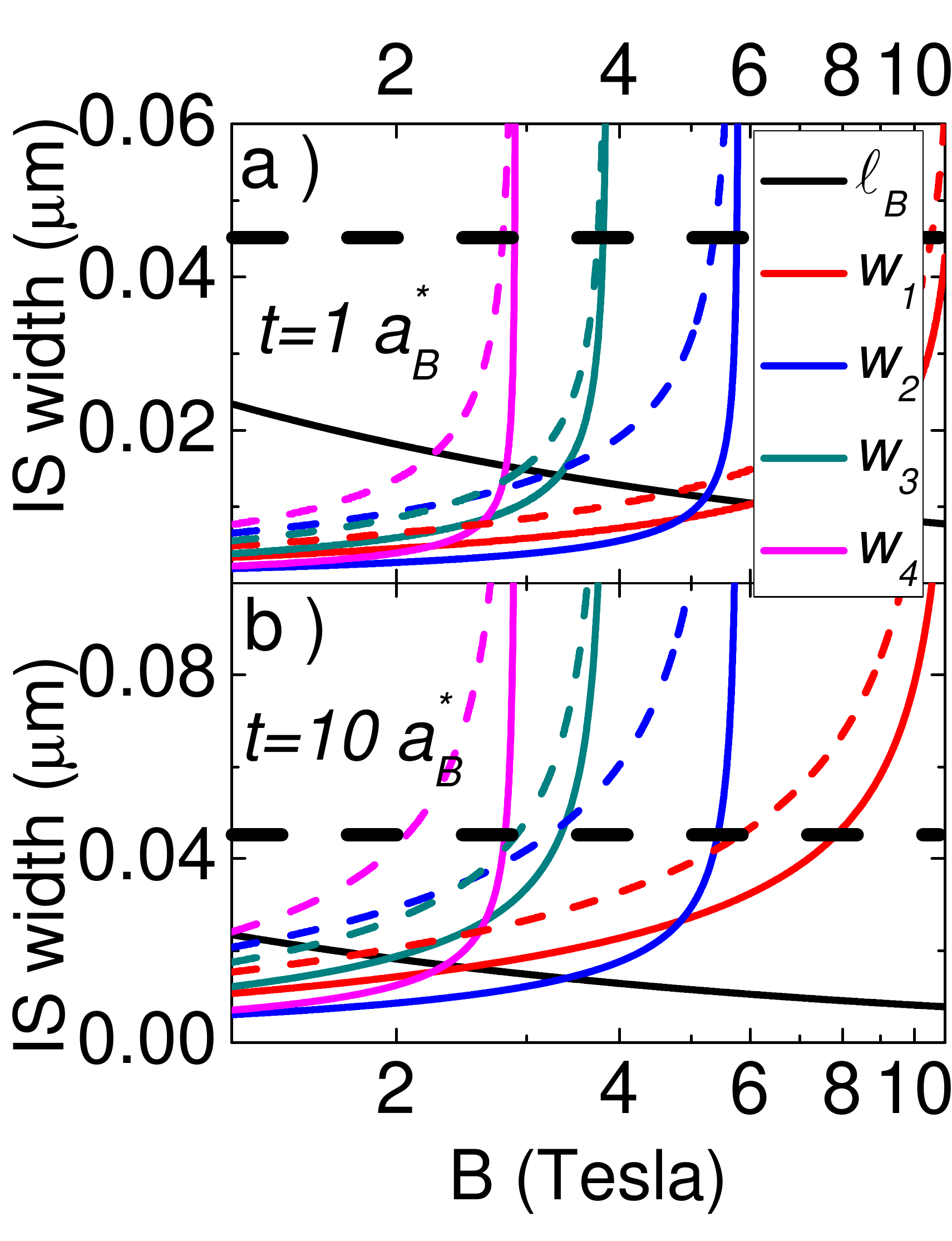}
\caption{\label{fig3} The widths of incompressible strips having four different filling factors as a function of magnetic field, calculated from Eq.~\ref{IS_width} using the self-consistent density profile, i.e Eq~\ref{afifhoca_density}. The solid lines denote $g^*=-0.44$, whereas broken lines depict $g^*=5.2$. (a) The steep edged QD, $t=a^*_B$. (b) The smooth edged QD, $t=10a^*_B$. In both cases depletion length is set to be $10a^*_B$.
}
\end{figure}
The spatial positions and the widths of the incompressible strips can be obtained analytically utilizing the formulation provided in Ref.~\cite{Chklovskii:1992}. The width of the $k^{\rm th}$ strip, where $k=\nu$ is an integer, is given by \cite{Chklovskii:1992}
\begin{equation}\label{IS_width}
W_{k}^2={\frac{2\kappa\Delta E}{\pi^2e^2 \frac{dn(r)}{dr }|_{r=r_k}}},
\end{equation}
here $\Delta E$ is the single particle energy gap, either due to Zeeman splitting (=$g^*\mu_BB$, with $\mu_B$ being the effective Bohr magneton and $\nu$ is an odd-integer) or due to Landau splitting (=$\hbar\omega_c-g^*\mu_BB$ and $\nu$ is an even-integer) and $\kappa~(=12.4)$ is the dielectric constant of the heterostructure.  Fig.~\ref{fig2}d shows the evolution of the strip widths as a function of external $B$ field, considering a chemically etched sample with bulk Land\'e $g^*$ factor. For all incompressible strips, one observes that their width decrease by decreasing $B$ field, with different rates depending on the filling fraction involved. The odd-integer strips become narrower faster than the even-integer strips, which is a consequence of the smaller energy gap of the Zeeman splitting compared to Landau splitting.

As mentioned previously, now we should check whether if the strip widths are larger than the thermodynamic length scale, $\lambda_T$, which is the Fermi wavelength at $T=0$. The horizontal broken line in Fig.~\ref{fig2}d depicts the Fermi wavelength, we can see that $\nu=1$ incompressible strip is larger than $\lambda_F$ for $B\gtrsim8$ T. Below this value, the strip becomes thermodynamically compressible. Let's consider the classical case, if the separation between plates $d$ becomes too small, such a capacitor will become leaky, while it is possible to have charge transfer from one plate to the other, regardless of the insulator in between. In such a situation diffusion takes place between the two surrounding compressible regions ($\nu<1$ and $\nu>1$), in thermodynamical terms. The same argument holds also for other filling factors at different field values. Therefore we should not threat incompressible strips as an insulator, since, when $W_k.\lambda_T\lesssim 1$, the strip is no longer incompressible thermodynamically \cite{Siddiki:2011,Wild:2010,Metin:2013}.

Another limit bounding the widths of the incompressible strips from below is the magnetic length, which is essentially the width of the wave function. In this case one can think of the incompressible strip as a barrier and once the wave functions of the compressible regions on each side of the barrier overlap, scattering takes place. Note that, only in the limit of $T\rightarrow0$ and without any disorder the scattering is zero, i.e. at any finite temperature and real sample with finite mobility, the scattering probability is finite due to finite TDOS at $E_F$. In Fig.~\ref{fig2}d, we also show the magnetic length as a function of $B$, which increases with $B$. As an example, for $\nu=1$ below 2.5 T it is possible to have tunnelling (or scattering) between the two compressible regions surrounding

From Eq.~\ref{IS_width}, one can clearly see that, the widths of the incompressible strips are essentially determined by the energy gap together with the strong density gradient dependency. Next, we investigate the effect of the edge steepness and the effect of effective $g^*$ on the magnetic field dependency of the incompressible strip widths. Fig.~\ref{fig3}, plots the widths of incompressible strips as a function of $B$, assuming two $g^*$ (solid lines, -0.44 and broken lines 5.2) and considering step edge, Fig.~\ref{fig3}a, and smooth edge, Fig.~\ref{fig3}b. The first observation is, when the Zeeman splitting is large the strips stay incompressible for larger $B$ intervals, which is more pronounced for the smooth edge profile. Meanwhile, if the edge is steeper then the the possibility to observe incompressible strips at the edges are suppressed. Interestingly, to observe two incompressible strips at the edge having different filling factors is not possible for edge profiles considered here. This implies that, only the inner strip is thermodynamically incompressible and is able to decouple the surrounding compressible regions. Here, the decoupling term is used both thermodynamically and electro-dynamically.

In the next discussion we will utilise the above electrostatic picture to obtain the geometric capacitances of a QD and determine quantum capacitance dominated $B$ field intervals also considering the spatial distribution of the incompressible strips.
\begin{figure}
\hspace{1.4cm}
\includegraphics[width=.9\columnwidth]{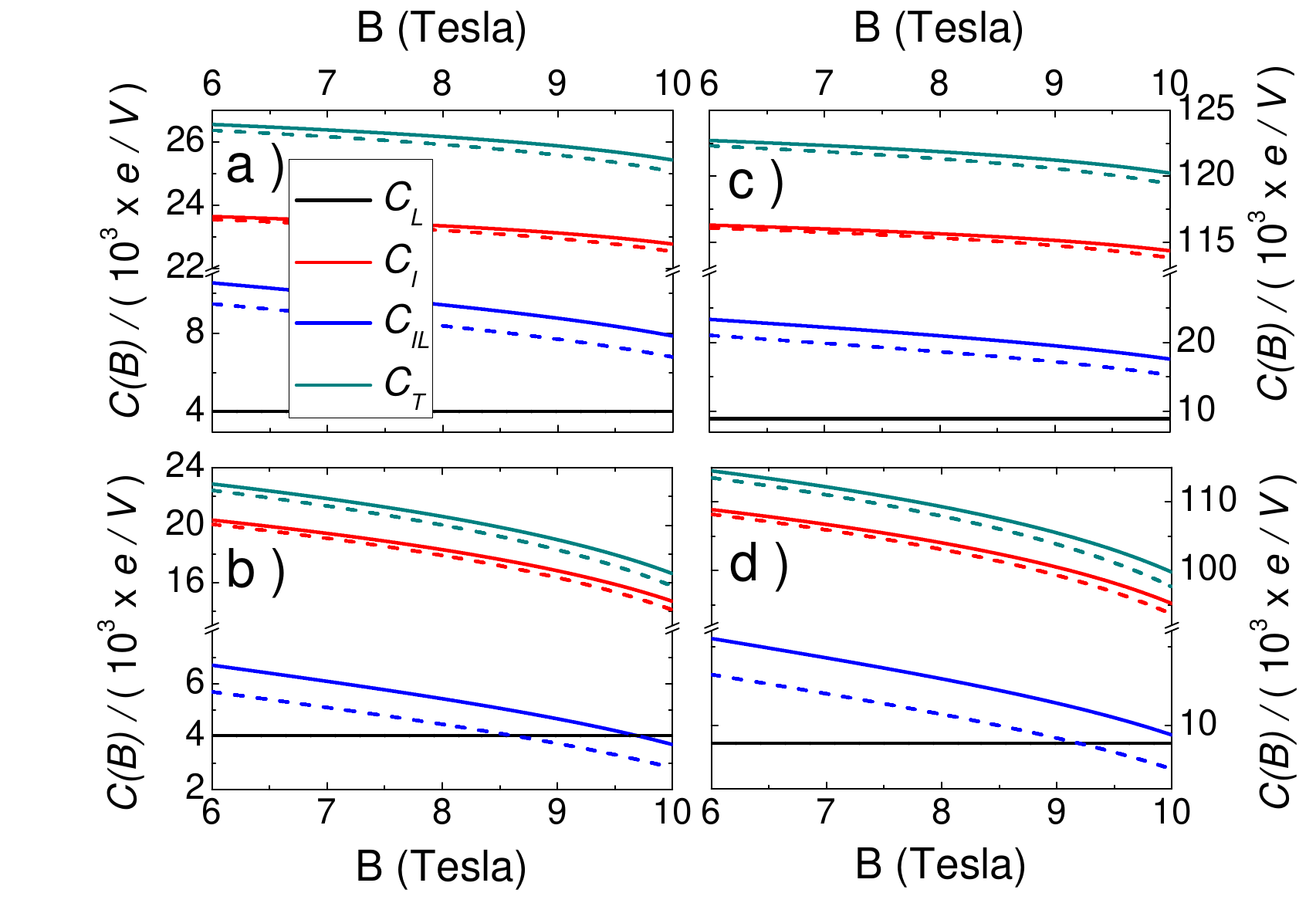}
\caption{ \label{fig4} Geometric capacitances of a small QD with a radius of $R=120a^*_B$ (a,b) and a large QD $R=250a^*_B$ (c,d). (a) and (c) demonstrates a steep edge, $t=a^*_B$, whereas (b), (d) depicts smooth edge, $t=10a^*_B$. In all cases solid lines denote $g^*=-0.44$ and broken lines denote $g^*=5.2$, with $\ell_d=10a^*_B$.}
\end{figure}
\subsubsection{Geometrical capacitances from electrostatics}
Once the widths of the insulating strips are obtained as a function of magnetic field, it is rather straight forward to calculate the geometrical capacitances. Here, we will only refer to the capacitances for similar structures proposed in the literature, namely to ones proposed by Evans et al and Halperin et al, to compare our findings with the existing results. For the sake of simplicity, we will first consider the two lowest Zeeman split Landau level, for which a single incompressible strip exists. Including higher levels result in capacitances in series and yield relatively complicated electric circuits, can be obtained with some tedious arithmetical calculations. However, as discussed, the outer most strips become transparent to radial electric field. Since, their widths become smaller than thermodynamic and the magnetic length at lower fields, hence, these insulating strips do not contribute to total geometric capacitance.

As a first step, we calculate the capacitance between the inner compressible region ($\nu>1$) and the outer compressible region $\nu<1$ separated by the $\nu=1$ incompressible strip, depicted as $C_{IL}$ in Fig.~\ref{fig2}a. Assuming that the width of the incompressible strip is narrow compared to the radius of the compressible region, one can determine the charge distribution in the close neighbourhood of the incompressible strip using two parallel conducting strips. Since the distance from the surface to the 2DES is larger than the widths of the incompressible strips then capacitance is given by \cite{Evans:1993},
\begin{equation}\label{Evans_C}
C_{IL}={\frac{\kappa L_{\rm IS}}{2\pi^2}}\ln(\frac{4d}{W_1}),
\end{equation}
where $L_{\rm IS}$ is the perimeter of the incompressible strip with $\nu=1$, $W_1$ is its width and $d$ is the distance between the top gate and the 2DES, taken to be 100 nm in accordance with common experimental structures. Similarly, the capacitance between the outer compressible strip with $\nu<1$ and the surrounding gate reads,
\begin{equation}\label{Evans_C1}
C_L={\frac{\kappa L_{\rm CS}}{2\pi^2}}\ln(\frac{4d}{\ell_d}),
\end{equation}
where $L_{\rm CS}$ denotes the outer perimeter of the compressible strip and remains unchanged with changing $B$, namely $L_{\rm CS}=2\pi R$. Note that the argument of the natural logarithm is also constant in $B$, since both the distance from the top gate and the depletion length are fixed. Finally, the capacitance between the top gate and the inner compressible region is given by
\begin{equation}\label{Evans_C2}
C_I={\kappa \frac{A_1}{d}}.
\end{equation}
Here, $A_1$ is the area of the inner metallic-like island with $\nu<1$. The total geometric capacitance of the equivalent electrical circuit shown in Fig.~\ref{fig2}a which is composed of capacitors described above reads,
\begin{equation}\label{Evans_CT}
C_{geo}=C_I+\frac{C_{IL}C_L}{(C_{IL}+C_L)},
\end{equation}
and is plotted in Fig.~\ref{fig4} together with $C_{IL}$, $C_L$ and $C_I$ as a function of $B$ considering various sample parameters. The upper and lower panels differ in edge steepness, whereas left and right panels show different sample sizes. In all plots the bulk effective $g^*$ is shown by solid lines, meanwhile the exchange enhanced effective $g^*$ is depicted by broken lines. The first observation is that the smaller sample has an order of magnitude smaller capacitance compared to large sample. The second observation is, the total geometric capacitance as well as $C_{IL}$ and $C_L$ increases by decreasing $B$ field. The $B$ dependency of the curves is mainly affected by the edge steepness rather than sample size. Capacitances of the smooth edge sample almost linearly increase with decreasing field and the steep edge samples follow the inverse behaviour of the strip widths as a function of the field. We should remind that, in the above discussion we considered a $B$ interval in which a single incompressible strip existed with $\nu=1$. However for lower fields, it might be possible to have co-existing incompressible strips with different filling factors, depending on the edge profile. In such a situation, one should follow simply the addition of serially connected capacitors to obtain related capacitances.



In the left panel of Fig.~\ref{fig5} we show the total geometric capacitances as a function of $B$ field, depicted by the solid line, calculated according to the addition rule for two different $g^*$. It is observed that, $C_{geo}$ increases stepwise for filling factors greater than $\nu=1$, where the amplitude depends on $g^*$.  The capacitance is constant for  $\nu<1$, reflecting constant area of the compressible island at the QD. For $g^*=-0.44$, both the even and odd integer filling factors with $\nu<5$ present stepwise increase of the capacitance. For higher filling factors stepwise increase is smeared out, while the energy gap becomes comparable with the thermal energy and incompressibility vanishes. At the larger $g^*$, the behaviour is repeated, however, since the Zeeman gap is relatively larger at $\nu=4$, we observe that two incompressible strips ($\nu=4$ and $\nu=3$) contribute to total capacitance and hence the geometric capacitance is decreased at $\nu=4$. Similar observation is also valid for other even integer filling factors, however, less pronounced.

So far we have only calculated the geometrical capacitances resulting from the co-existence of compressible and incompressible strips. However, as mentioned in the Introduction, one can also define a capacitance that depends on the density of states of the system. Since, to add a particle to a system with finite DOS requires some work to be done on the system. Notably, for 2DES without an external $B$ field the DOS is constant $D_0=m^*/\pi\hbar^2$, hence, the number of particles to be added is constrained by the area of the system. However, in the presence of an external field the DOS varies as a function of the field strength yielding a field dependent capacitance. In the next subsection we will calculate this quantum capacitance and investigate its influence on charging energy, later. Fig.~\ref{fig5}, left panel depicts the total quantum capacitances by broken lines together with the total capacitance composed of both geometric and quantum counter parts. We observe that, $C_q$ vanishes at integer filling factors for $\nu<4$ and is at least an order of magnitude smaller compared to $C_{geo}$, hence, the total capacitance is dominated by the quantum counter part. In particular, when there exists an incompressible strip both the total and the quantum capacitances vanish. In between the integer filling factors, the geometric capacitance also contributes to the total capacitance.

In the next section we will investigate the effect of the total capacitance on Aharonov-Bohm interference induced conductance oscillations following the pioneering work of Beenakker et. al.

\section{Coulomb blockade of Aharonov-Bohm oscillations }
\begin{figure}
\hspace{1.7cm}
\includegraphics[width=.94\columnwidth]{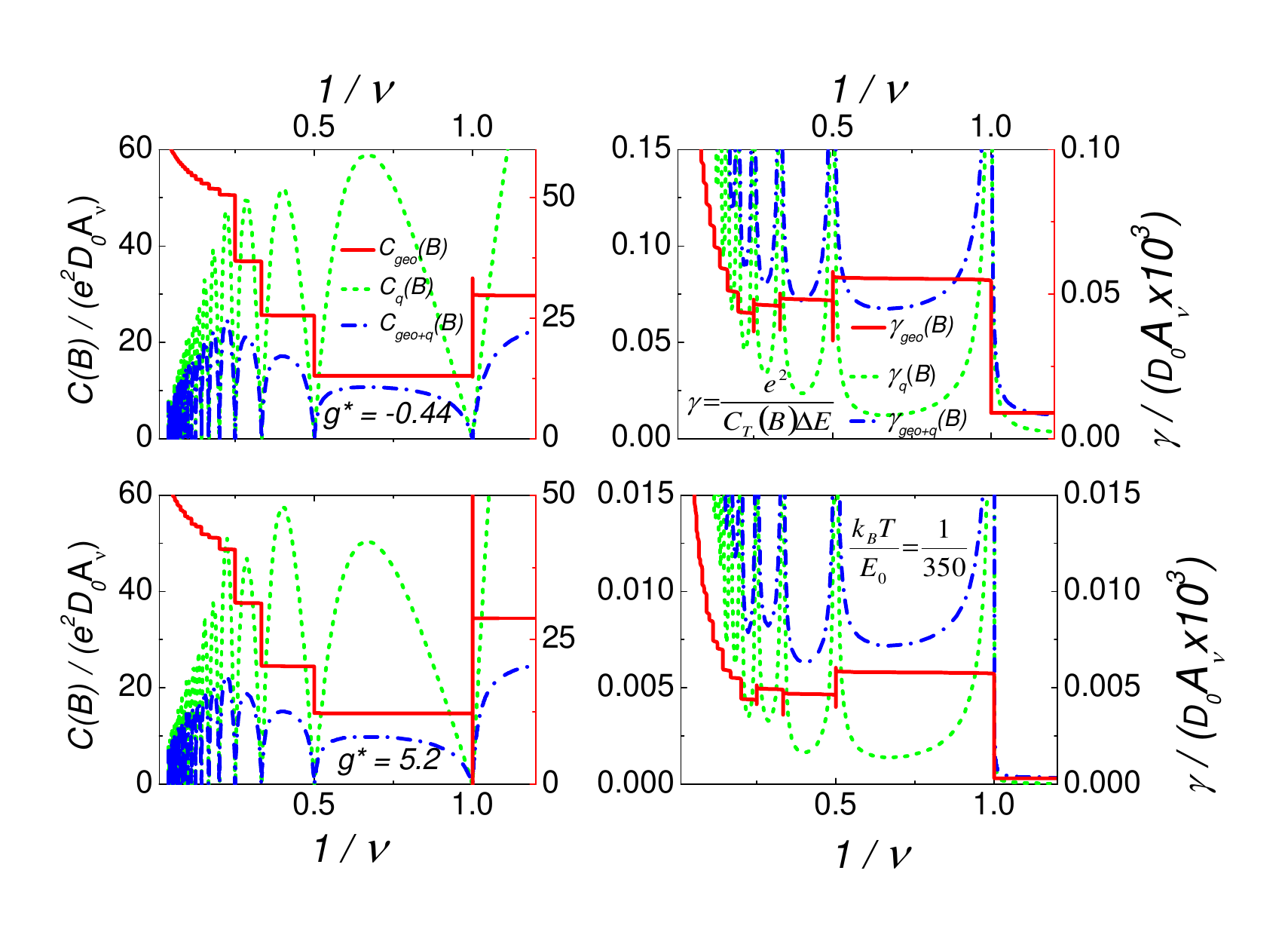}
\caption{ \label{fig5} (a, b, c) Geometric, quantum and total capacitances of a small QD with a radius of $R=120a^*_B$ without (a) and with spin splitting, i.e. $g^*=-0.44$ (b) and $g^*=5.2$ (c). (d,e, f) Calculated phase corrections $\gamma$ according to Eq.~\ref{Magnetoconductance_star}.}
\end{figure}

More than three decades ago, Coulomb blockade of the Aharonov-Bohm oscillations were predicted for a QD operating under integer quantized Hall conditions \cite{Beenakker:1991}. There, using energy arguments and the electrostatic stability conditions it is shown that the charging energy of the QD enhances the periodicity, when conductance is measured as a function of the $B$ field. In the original AB interference experiments, the field does not penetrate the path of the electrons, hence the $\Delta B$ periodicity is unaffected by the interactions. However, in our case the flux also exists where the interfering electrons reside. Therefore, it is expected that the $\Delta B$ period should be also effected by interactions, namely by the charging energy \be E_C= e^2/C. \ee
Given the charging energy, the energy level spacing $\Delta E$ is renormalised and can be expressed as $\Delta E^*=\Delta E+E_C$ and one can write the renormalised periodicity,
\begin{equation}\label{Magnetoconductance_star}
\Delta B^*=\Delta B(1+\gamma),
\end{equation}
where phase shift $\gamma$ equals $=e^2/{C\Delta E}$ and unperturbed periodicity is $\Delta B=h/eA$, $A$ being the enclosed area of the QD. In Fig.~\ref{fig5}b and Fig.~\ref{fig5}d, we show the calculated phase shifts from the geometric capacitances and also taking into account the quantum capacitances. The phase shift, when only the geometric capacitance is considered is at the order of $10^{-3}$ for $g^*=5.2$, meanwhile is at most $6 \%$ for bulk $g^*$ \cite{Rosenow:2009}. Hence, is not able to explain shifts reported at the recent experiments, while in experiments the shifts are at the order of unity. In the next step, we also calculated $\gamma_{geo+q}$, the shift obtained by considering both the geometric and quantum capacitances, and show in the same figure. We observe that, the quantum capacitance increases $\gamma_{geo+q}$ at integer filling factors dramatically and the phase shift becomes at the order of unity.

At a first glance, one can think that taking into account quantum capacitances can explain the experimental findings. However, note that order of unity phase shifts only occur at a very limited $B$ field interval, in contrast, experimental findings report that such oscillations can be observed throughout the inter plateau interval. In addition, most of the experiments report AB oscillations at out of the plateau (to be specific, at the lower field part) intervals, where $\gamma_{geo+q}$ is still less than $10 \%$ for $g^*=5.2$. Finally, only taking into account the phase shifts can not explain the unexpected periodicity in $V_g$.

In the next Section, we will utilise the formulation developed by Evans et al to investigate the effect of quantum capacitance on the gate voltage periodicity \cite{Evans:1993}.

\begin{figure}
\hspace{1.7cm}
\includegraphics[width=.93\columnwidth]{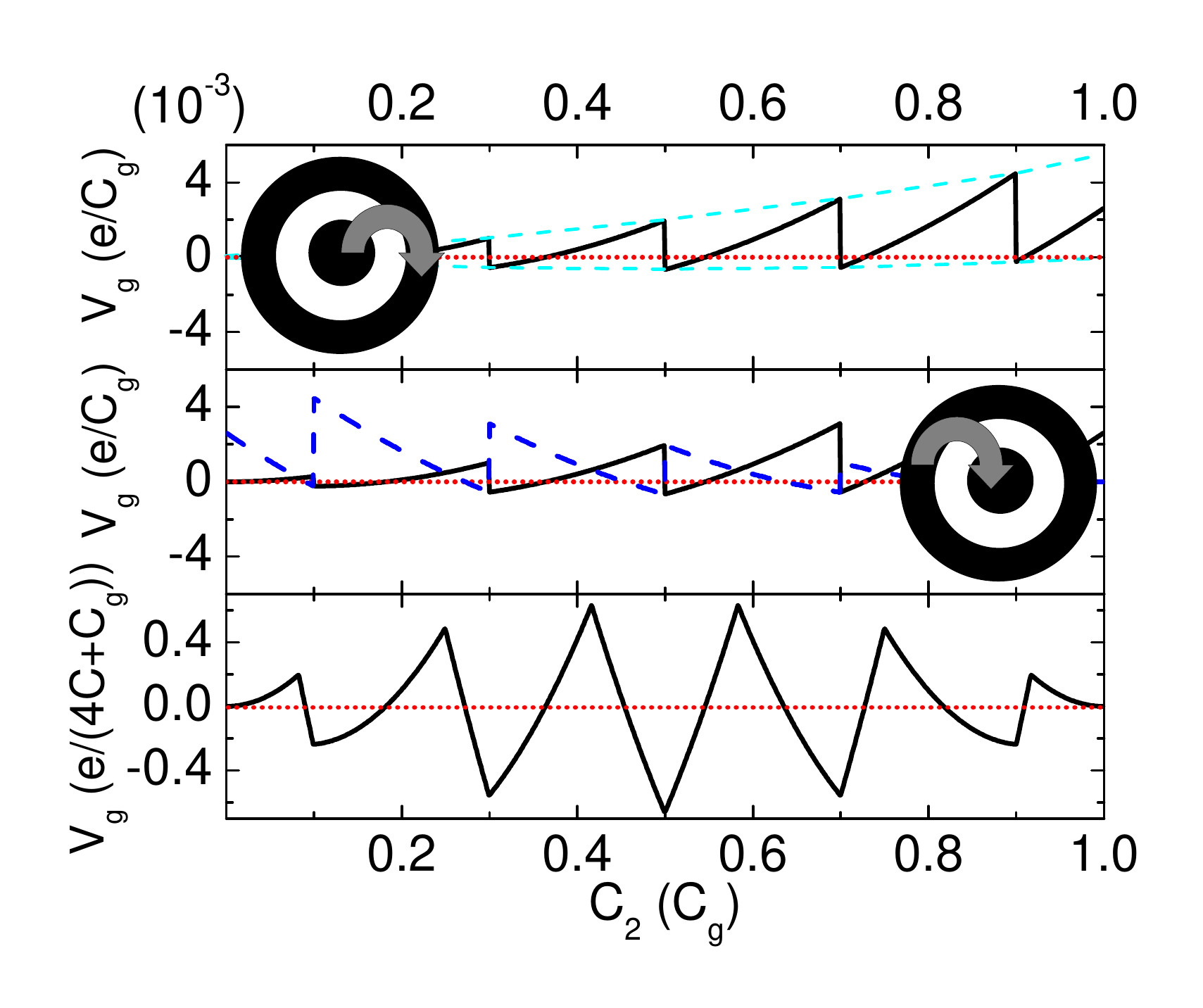}
\caption{ \label{fig6} The capacitance dependent charging voltage $V_g(C_2)$ calculated using the density profile given in Eq.~\ref{afifhoca_density} and incompressible strip width defined by Eq.~\ref{IS_width}. (a) In the case when the outer conductor is charged. (b) Same as (a) while charging the inner conductor. (c) Physically observable conductance oscillations due to charging effects.
}
\end{figure}
\begin{figure}
\hspace{1.7cm}
\includegraphics[width=.93\columnwidth]{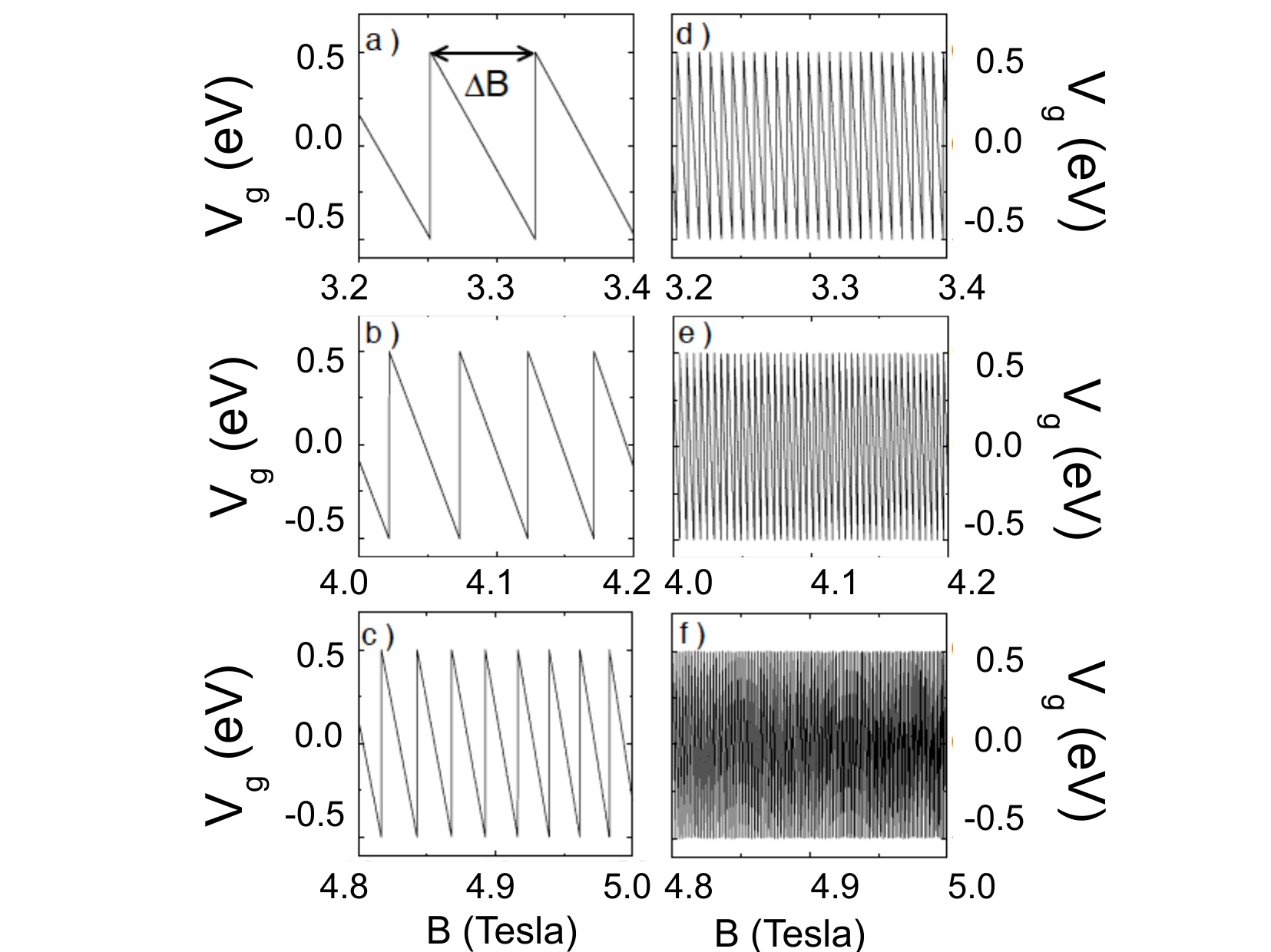}
\caption{ \label{fig7} The conductance oscillations as a function of the field. Left panel (a,b,c) shows the oscillation period considering a small smooth edge ($t=a^*_B$) sample, whereas right panel depicts the same quantity for a steep edge ($t=10a^*_B$) sample, at different filling factors.}
\end{figure}
\begin{figure}
\hspace{2cm}
\includegraphics[width=.89\columnwidth]{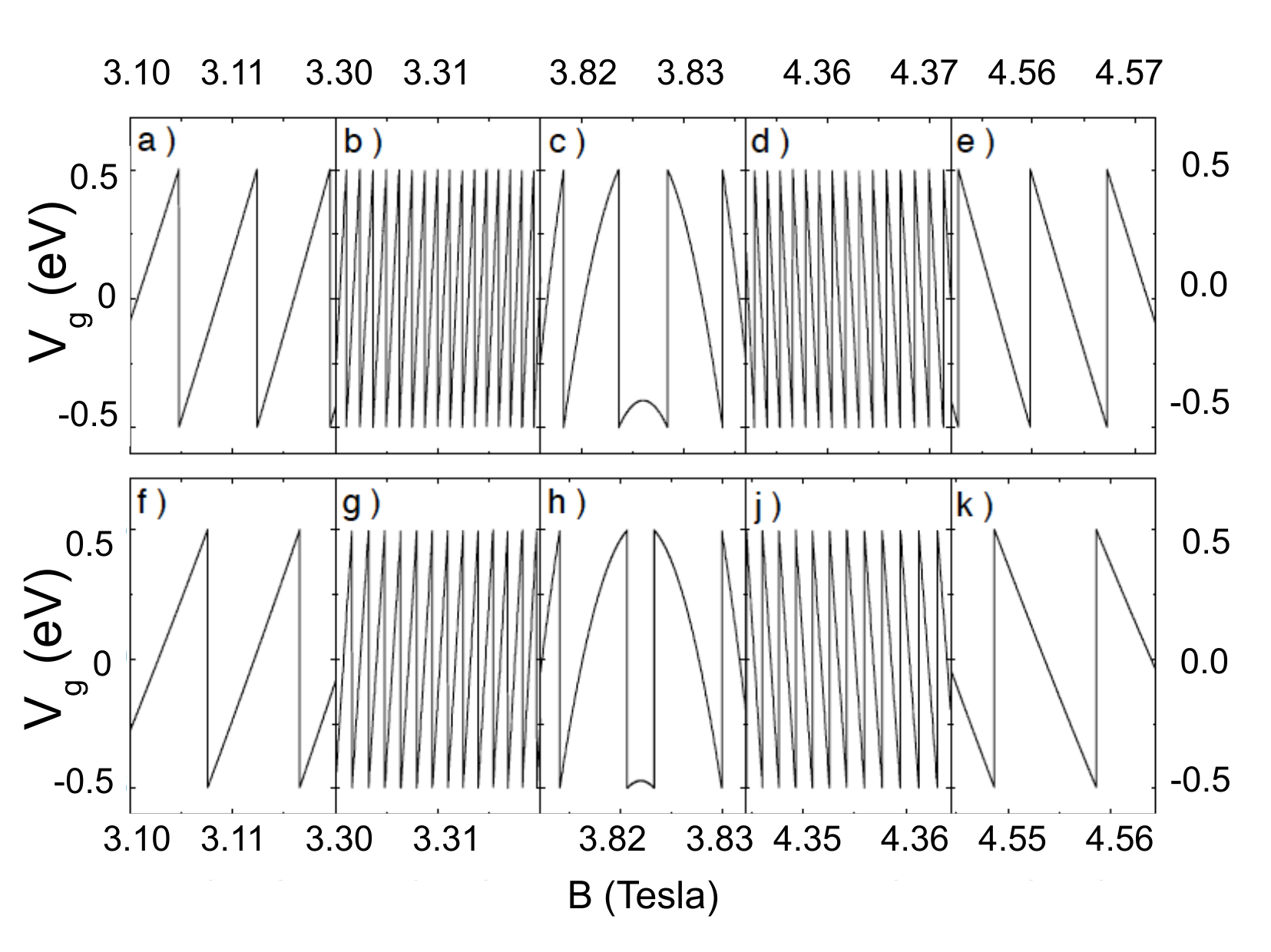}
\caption{ \label{fig8} The conductance oscillations as a function of the field, where quantum capacitance is also taken into account considering a small sample. At the upper panel, simulation results considering a smooth edged dot is shown, whereas lower panel depicts the same quantity for a steep edged dot. It is seen that the steepness has an observable difference between two edge profiles at both samples. This already points the fact that at small samples steepness is effective in determining oscillation periods. In accordance with experimental findings.
}
\end{figure}

\begin{figure}
\hspace{2.5cm}
\includegraphics[width=.8\columnwidth]{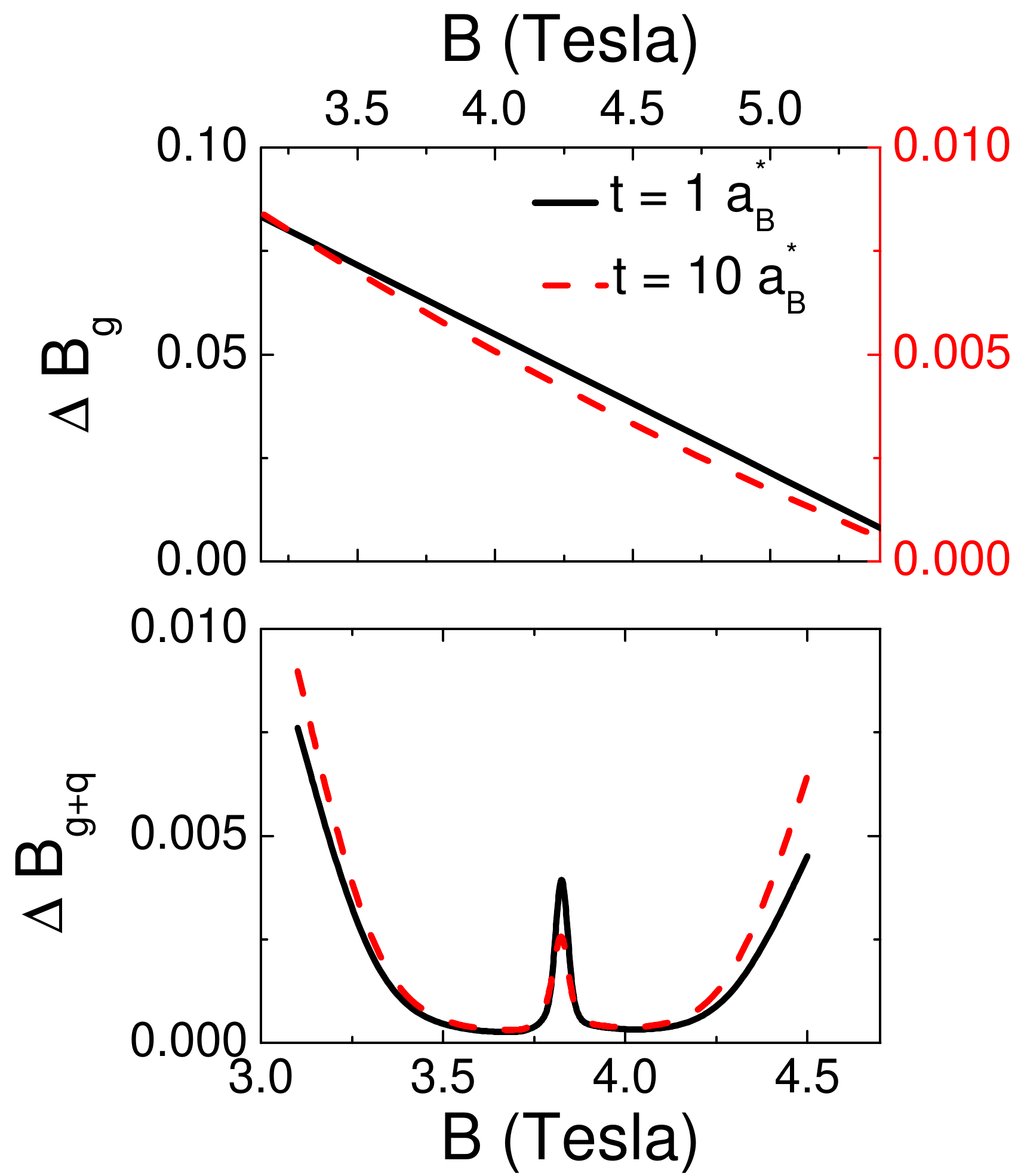}
\caption{ \label{katman_sonuclar_gateler} The field dependency of the magnetic field periodicity, depending on the edge potential profile. (a) Only considering the geometric capacitances. (b) Also including quantum capacitances.
}
\end{figure}

\section{CONDUCTANCE PEAK STRUCTURE DERIVED FROM THE ELECTROSTATIC MODEL \label{sec:cap_halperin}}
Here, we utilise the previously calculated capacitances of the QD shown in Fig. \ref{fig2}a and obtain the periodicity of the Coulomb Blockade oscillations considering a model including only three mutual capacitances as a function of the gate potential. The capacitance $C=C_{IL}$ is between two conducting regions, and $C_{1}=C_{I}$, $C_{2}=C_{L}$ are the capacitances to the gate. The details of the model can be found in the original work by Evans et al., however, our model differs from the original work where the $B$ field dependency of the capacitances even for the geometric case is neglected. Here, we explicitly calculate the $B$ field dependency of the incompressible strip widths and also include TDOS enriched total capacitance.

Now we briefly summarise the electrostatic argumentation of the original work. At $B=0$, there is no insulating strip, hence, if the capacitances of the compressible electron island to the leads can be neglected, the electrostatic energy is given by \cite{Evans:1993}
\begin{equation}\label{Electrostatic_energy}
U_{0}(N)=\frac{(Ne)^{2}}{2C_{g}}-{NeV_{g}},
\end{equation}
where, $C_{g}(=C_{1}+C_{2})$ is a constant, since the area of dot is independent of the magnetic field. Here, $N$ is the total number of the electrons in the system, and $V_{g}$ is potential difference between source and drain leads. In the presence of an external field incompressible strips form and transferring an electron from the inner compressible region to outer compressible region requires extra energy, at low temperatures and small bias voltage $V_{g}$ this energy is
\begin{equation}\label{Energy_different}
U(N+1)-U(N)=E_{F},
\end{equation}
where $E_{F}$ is the Fermi energy and $U$ is the electrostatic energy described by,

\begin{equation}\label{Polarization_charges_with_energy}
U=\frac{p^{2}}{2C_{1}}+\frac{q^{2}}{2C_{2}}+\frac{r^{2}}{2C}+\frac{s^{2}}{2C_{LD}}-V_{g}(p+q),
\end{equation}
here  $p,q,r$ and $s$ are the polarisation charges given as,
\begin{equation}\label{eq_a}
en_{1}=p+s-r,
\end{equation}
\begin{equation}\label{eq_b}
en_{2}=q+r,
\end{equation}
\begin{equation}\label{eq_c}
V_{g}=\frac{p}{C_{1}}-\frac{s}{C_{LD}}
\end{equation}
and
\begin{equation}\label{eq_d}
V_{g}=\frac{q}{C_{2}}-\frac{r}{C}-\frac{s}{C_{LD}},
\end{equation}
here, $n_{1}$ is electron number in the outer conducting area, $n_{2}$ is electron number in the inner conducting area and total electron number is $N=n_{1}+n_{2}$. Using these four equations one can obtain the polarisation charges from the Kirchhoff law, energy and charge conservation and finally express the $B$ field dependent conductance oscillations. We assume that the lead capacitances are zero, hence polarisation charge $s$ also vanishes. First, to obtain the energy of system, we determine the polarisation charges as a function of $n_{1}$ and $n_{2}$, using Eq. ({\ref{eq_a}}-{\ref{eq_d}}),
\begin{equation}\label{pol_p}
p=\frac{eCC_{1}(n_{1}+n_{2})+eC_{1}C_{2}n_{1}}{CC_{1}+CC_{2}+C_{1}C_{2}},
\end{equation}
\begin{equation}\label{pol_q}
q=\frac{eC_{2}C(n_{1}+n_{2})+eC_{1}C_{2}n_{2}}{CC_{1}+CC_{2}+C_{1}C_{2}},
\end{equation}
and
\begin{equation}\label{pol_r}
r=\frac{eC(C_{1}n_{2}-C_{2}n_{1})}{CC_{1}+CC_{2}+C_{1}C_{2}}.
\end{equation}
In the second step, we obtain the total electrostatic energy replacing the above calculated polarisation charges in Eq. (\ref{Polarization_charges_with_energy}),

\begin{equation}\label{total_electrostatic_energy}
U=-eV_{g}N+\frac{e^{2}N^{2}}{2C_{g}}+ \frac{e^{2}(n_{1}C_{2}-n_{2}C_{1})^{2}}{2C_{g}(CC_{g}+C_{1}C_{2})}.
\end{equation}

In the case of zero magnetic field, we can consider the QD as a single conducting region, corresponding to the limit $C\rightarrow\infty$ and Eq. (\ref{Polarization_charges_with_energy}) then reduces to Eq. (\ref{total_electrostatic_energy}). The first two terms in Eq. (\ref{total_electrostatic_energy}) are independent of the charge distribution between the two conductors and depend only on $N$. The third term therefore determines the charge distribution inside the dot. The equilibrium value of $n_{1}$ can be found by minimising the third term in Eq. (\ref{total_electrostatic_energy}) with respect to variations in $n_{1}$, holding $N$ fixed,
\begin{equation}\label{n1_ust}
n'_{1}=\frac{C_{1}N}{C_{g}},
\end{equation}
$n'_{2}$' can also be determined similarly. Rewriting Eq. (\ref{Electrostatic_energy}) using $n'_{1}$ and $n'_{2}$ and terms  we obtain the explicit form of the $V_g$ as a function of capacitances, which now also depends implicitly on $B$,
\begin{equation}\label{Vg_oscillations}
V_{g}=-\frac{E_{F}}{e}+\frac{(2N+1)e}{2C_{g}}+ \frac{eC^{2}_{2}}{2C_{g}(C_{g}C+C_{1}C_{2})}+\frac{eC_{2}(n'_{2}-n_{2})}{C_{g}C+C_{1}C_{2}}
\end{equation}

For the sake of consistency, we will first repeat the calculation of Evans et al., however, in our calculations we will include the $B$ field dependency of the incompressible strip widths and the area of the compressible regions obtained as previously. Fig.~\ref{fig6} plots the capacitance dependency of the voltage (conductance) oscillations, in Fig.~\ref{fig6}a the case where the an electron is brought to inner conductor, whereas in Fig.~\ref{fig6}b an electron is placed at the outer conductor is shown. The total voltage oscillations are depicted in Fig.~\ref{fig6}c. For comparison we refer to Fig.4, where they constructed the voltage oscillations by assuming $C_2$ is a monotonous function of the $B$ field. We observe that, our self-consistent calculations that also take into account the widths of the incompressible strips modify their picture slightly. Namely, the conductance oscillations are no longer triangular shaped and sharp, but smoothened due to $B$ field dependency.

Next, we investigate the conductance oscillations as a function of $B$ field considering two different edge profiles for a small QD ($R=120a^*_B$), where only the geometric capacitances are taken into account. The left panel of Fig.~\ref{fig7} plots the conductance oscillations at different magnetic field intervals for a smooth edged QD, where we observe that the period $\Delta B$ strongly depends on $B$. For the steep edge sample (right panel, Fig.~\ref{fig7}), the period is an order of magnitude smaller and is also strongly $B$ dependent. These two behaviours can be understood quite easily: For the smooth edged sample the incompressible strips reside closer to the bulk of the QD and the area covered by the compressible region in the bulk is small. For the steep edged sample, the opposite behaviour is expected where incompressible strips reside closer to the edges and area is larger. As a direct consequence of this simple areal dependency the $\Delta B$ would be large for the smooth edged sample compared to steep edged. However, they will present similar $B$ field dependency.  Comparing Fig.~\ref{fig7} with Fig.~\ref{fig6}, we can conclude that by changing the field one only transfers charge from outer compressible strip to the inner compressible area

Now we are in a position to include quantum capacitances to our calculations. Fig.~\ref{fig8}, presents the same quantity as shown in Fig.~\ref{fig7}, however, here we also include the quantum capacitance to our calculations. The effect of edge steepness is not as pronounced strongly, since the quantum capacitance is now the dominating parameter. The geometric capacitance only depends on the areas of compressible and incompressible regions (or strips). In contrast, the quantum capacitance is mainly determined by the amount of available TDOS, which essentially determines the oscillation period. Remarkably, at the geometric case the slope of $V_g/B$ is always same, whereas once the quantum capacitance is taken into account this slope changes sign in proximity of bulk integer filling factors. To investigate the magnetic field periodicity $\Delta B$ as a function of $B$ we calculated the change of $\Delta B$ considering both different edge profiles and considering only geometric and both geometric and quantum capacitances, Fig.\ref{katman_sonuclar_gateler}. For the geometric capacitance, the period $\Delta B$ decreases linearly with increasing field. Interestingly the value of periods differ as much as an order of magnitude. Meanwhile, once the quantum capacitance is taken into account the
regardless of the steepness the period is at the same order of magnitude. As mentioned previously, while approaching an incompressible bulk ($B\sim 3.8$ T)  both from left (low $B$) or right (high $B$) the value of $\Delta B$ decreases in a non-linear manner. In the close proximity of integer bulk the period remains approximately unchanged. However, once the very	centre of the dot becomes compressible once more with a smaller filling factor, the period increases rapidly until the total area of compressible and incompressible strips equate. Clearly stating, if the TDOS (which depends on area) of incompressible and compressible areas contribute quantum capacitance equally the phase shift in $\Delta B$ saturates. For higher fields the compressible are at the centre increases and leads to a lower charging energy, decreasing the $\Delta B$ period until the incompressible strip surrounding the central compressible region becomes larger in area.

\section{CONCLUSION}

The long standing debate on the explanation of the observed conduction oscillations at 2D electron systems subject to high perpendicular fields is tackled by many outstanding physicists. It is claimed within the main stream approach that the observed oscillations are due to charging effects, whereas other possible mechanisms are also provided, as mentioned in Introduction. Here, we investigated the limitations of charging effect explanation also considering a reasonable model by handling the capacitive coupling proposition via calculating both the geometric and quantum capacitances also taking into account experimental device properties. We found that, without taking into account quantum capacitance, which is determined by TDOS, and the experimental properties of the device it is barely possible to provide a realistic and comprehensive explanation of experimental findings, utilising naive single particle edge state picture.

Our approach is promising in couple of senses: First we can handle the classical electrostatic charging problem of a quantum dot subject to perpendicular magnetic fields in a self consistent manner by calculating the widths and positions of incompressible (insulating) strips. Second, we calculated quantum capacitance of the device considering the existence of compressible and incompressible regions as a function of magnetic field and sample properties, such as area and edge profile of the devices. Up to our knowledge, no similar approach has been reported at the literature. On one hand, our findings reveal many of the unexplained observations, such as the variation of the $\Delta B$ period, including the constant regime, and the effect of geometric and lithographic properties of the sample. Our results show that, the commonly accepted theory of the Fabry-P\'{e}rot interferometers may become questionable once real experimental conditions are taken into account. On the other hand, in experiments different size qdots are defined by electrostatic means, e.g. directly gate \cite{marcus:01} or trench gating \cite{Goldman:2005}, hence we claim that our formulation can explain both experimental geometries and also the cross-over between them.

\section*{Acknowledgements}
We would like to express our gratitude to B. Halperin, M. Heiblum, B. Rosenow, K. von Klitzing, E. J Heller and R. R. Gerhardts both for their fruitful discussions and sincere criticism. This work is financially supported by the scientific and research council of Turkey (T\"UB\.ITAK) under grant no: TBAG-112T264 and 211T264.

\section*{References}

\bibliographystyle{elsarticle-num}

\end{document}